\documentclass{aa}

\usepackage{graphicx}
\usepackage{txfonts}
\usepackage{booktabs}

\usepackage{dcolumn}
\usepackage{bm}
\usepackage{color}
\usepackage{hyperref}

\usepackage{amsmath}

\usepackage{natbib}
\bibpunct{(}{)}{;}{a}{}{,}
\usepackage{orcidlink}

\begin{document}

    \title{The Big Sobol Sequence: how many simulations do we need for simulation-based inference in cosmology?}

    \subtitle{}

    \author{Anirban Bairagi
          \inst{1}\,\orcidlink{0009-0009-3089-052X} 
          \and
          Benjamin Wandelt\inst{1,2,3,4}\,\orcidlink{0000-0002-5854-8269}
          \and
          Francisco Villaescusa-Navarro\inst{4,5}\,\orcidlink{0000-0002-4816-0455}
            }
   \institute{%
CNRS \& Sorbonne Universit\'{e}, Institut d’Astrophysique de Paris (IAP), UMR 7095, 98 bis bd Arago, F-75014 Paris, France\;\;\\
              \email{bairagi@iap.fr}         
        \and
            Department of Physics and Astronomy, Johns Hopkins University, 3400 North Charles Street, Baltimore, MD 21218, USA 
        \and
            Department of Applied Mathematics and Statistics, Johns Hopkins University, 3400 North Charles Street, Baltimore, MD 21218, USA 
        \and
             Center for Computational Astrophysics, Flatiron Institute, 162 5th Avenue, New York, NY 10010, USA 
        \and
            Department of Astrophysical Sciences, Princeton University, Peyton Hall, Princeton, NJ 08544-0010, USA
             }

    \abstract
{How many simulations do we need to train machine learning methods to extract information available from summary statistics of the cosmological density field? Neural methods have shown the potential to extract non-linear information available from cosmological data. To achieve this depends on having appropriate network architectures and a sufficient number of simulations for training the networks. In the first detailed convergence study of neural network (NN) training to extract maximally informative summary statistics for cosmological inference, we show that currently available simulation suites, such as the Quijote Latin Hypercube(LH) with 2000 simulations, do not provide sufficient training data for a generic neural network to reach the optimal regime. We present a case study in the context of training a moment network to infer cosmological parameters from the nonlinear dark matter power spectrum, where the optimal information content can be computed through asymptotic analysis using the Cramér-Rao information bound. We find an empirical neural scaling law that predicts how much information a neural network can extract from highly informative summary statistics as a function of the number of simulations used to train the network, for a wide range of architectures and hyperparameters. Looking beyond two-point statistics, we find a similar scaling law for training neural posterior inference using wavelet scattering transform coefficients. To verify our method, we created the largest publicly released suite of cosmological simulations, the Big Sobol Sequence(BSQ), consisting of 32,768 $\Lambda$CDM n-body simulations uniformly covering the $\Lambda$CDM parameter space. Our method enables efficient planning of simulation campaigns for machine learning applications in cosmology, while the BSQ dataset provides an unprecedented resource for studying the convergence behavior of neural networks in cosmological parameter inference. Our results suggest that new large simulation suites or new training approaches will be necessary to achieve information-optimal parameter inference from non-linear simulations.}

    \titlerunning{Neural Scaling laws}
    \authorrunning{A. Bairagi et al.}
    \maketitle

\section{\label{sec: intro}Introduction\protect\\}
A major effort in modern cosmology is directed towards inferring cosmological parameters and testing the fundamental tenets of the standard model, $\Lambda$CDM \citep{Ostriker:1995su, Planck:2013pxb}, which posits that the universe's energy budget is dominated by a cosmological constant ($\Lambda$) representing dark energy ($\sim$70\%) and dark matter ($\sim$25\%). This endeavor is supported both by numerical simulations and forthcoming data from a new generation of major cosmological surveys. These include large-scale structure surveys such as the Dark Energy Spectroscopic Instrument (DESI) \citep{DESI_Collaboration_2022}, the Vera C. Rubin Observatory \citep{LSST:2009pqm}, Euclid \citep{Laureijs:2011gra}, the Nancy Grace Roman Space Telescope \citep{spergel2015widefieldinfrarredsurveytelescopeastrophysics}
, and SPHEREx \citep{SPHEREx:2014bgr}, as well as cosmic microwave background experiments like the Simons Observatory \citep{Ade_2019} and CMB-S4 \citep{abazajian2016cmbs4sciencebookedition}.

While sophisticated methods for cosmological parameter inference have dramatically reduced our uncertainty about cosmological parameters from past surveys, e.g., WMAP \citep{2003ApJS..148....1B}, Planck \citep{PlanckOverview2020}, DES \citep{2022PhRvD.105b3520A}, BOSS \citep{BOSS:2012dmf, Alam_2017}, DESI \citep{DESI:2024jis} in recent decades, current and upcoming surveys present new methodological challenges. The traditional approach compares summary statistics \citep{peeblesLSS, Peebles:1980yev}---such as n-point correlation functions \citep{peebles2001galaxymassnpointcorrelation, Scoccimarro:2000sn, doi:10.1073/pnas.2111366119}, number counts \citep{artis2024srgerositaallskysurvey} from various probes, i.e., galaxy clustering \citep{1977SciAm.237e..76G, Kaiser:1987qv} and weak lensing \citep{Heymans:2013fya, Heymans:2020gsg, Liu_2015}---against theoretical predictions or numerical simulations. These methods typically assume an explicit form of the likelihood function, often Gaussian, but this assumption becomes problematic in the non-linear regime where the true likelihood form may not be known, especially in the regime of a few modes and complex modeling. For example, even when the number of modes is large, non-linear evolution can lead to strong correlations, slowing the asymptotic approach to the Gaussian form \citep{Sellentin_2017, Hahn_2019}. Even when good approximations do exist, standard Bayesian inference techniques like Markov Chain Monte Carlo (MCMC) can face convergence challenges \citep{LiddleStatisticalMethodsReview2009, Leclercq:2014jda}.

Implicit Inference (II), or Simulation-Based Inference (SBI), has emerged as a promising alternative that learns the joint statistics of data and parameters directly from simulated pairs, avoiding the need for explicit likelihood assumptions. SBI has demonstrated both accuracy and computational efficiency across various applications \citep{Alsing_2018, Alsing_2019, Cranmer_2020, jeffrey2020solvinghighdimensionalparameterinference, KIDS-1000,coleFastCredibleLikelihoodFree2022, Lemos_2023}. Neural networks play two crucial roles in modern SBI approaches: generating or compressing informative summary statistics \citep{jeffrey2020solvinghighdimensionalparameterinference, lemos2023simbigfieldlevelsimulationbasedinference, Novaes:2024dyh} and representing statistical distributions such as the likelihood or posterior. Information Maximizing Neural Networks (IMNN) exemplify the first role, focusing on Fisher information maximization \citep{Charnock_2018}, while (ensembles of) normalizing flows \citep{papamakarios2021normalizingflowsprobabilisticmodeling, papamakarios2018maskedautoregressiveflowdensity}, neural ratio estimation \citep{coleFastCredibleLikelihoodFree2022}, and diffusion models \citep{song2021scorebasedgenerativemodelingstochastic, ho2020denoisingdiffusionprobabilisticmodels, song2022denoisingdiffusionimplicitmodels} are examples of the second.

While both of these tasks will require assessment of convergence, in this paper we will focus on the question of information extraction\footnote{For a recent paper addressing the convergence of neural density estimators in the cosmological context, see \cite{homer2024simulationbasedinferencedodelsonschneidereffect}.}, for the following reasons: 1) suboptimal information extraction will affect downstream tasks such as the inference of the likelihood or the posterior distribution; and 2) it is well known that the difficulty of fitting general high-dimensional distributions scales poorly with the number of dimensions, highlighting the utility of extracting information in a low-dimensional feature vector. Even in cases where neural density estimators are trained conditional on the feature vector, it has been observed in several of the above-cited works that providing long, uncompressed data vectors to conditional neural density estimators leads to suboptimal performance.

Given the centrality of trained neural networks for the extraction of informative features, a key question emerges: How many simulations are needed to achieve near-optimal results? While one could generate simulations until the extracted information plateaus, this brute-force approach is computationally expensive or even intractable. While even suboptimal information extraction may be an improvement over traditional summary statistics, it is clearly interesting to have a simple tool based on a power law fit to assess whether the extracted information from valuable data could be improved by running more training simulations. Our work presents the first comprehensive analysis of neural network convergence in cosmological information extraction and for relatively low-dimensional summary statistics such as the power spectrum provides a predictive framework to determine the required number of simulations based on numerical evaluation of the Fisher information.

For our investigation, we focus on the non-linear dark matter power spectrum $P(k)$ as our main summary statistic. This choice is motivated by the fundamental role of $P(k)$ in cosmology: at large scales and high redshifts, where the universe remains approximately Gaussian and homogeneous, $P(k)$ captures the complete statistical description \citep{Kaiser:1987qv, Peacock:1996ci, Gil_Mar_n_2012, Vlah_2016}. While current three-dimensional surveys access significant non-Gaussian information at smaller scales and lower redshifts, any method claiming to extract cosmological information from the non-linear regime should, at minimum, demonstrate the ability to do so with the power spectrum \citep{SDSS:2003tbn, SDSS:2004kqt, PhysRevD.105.043517, DAmico:2022osl}. Our findings reveal that even for this basic case, the required training set size is substantial, suggesting that simulation requirements may limit the amount of information extracted from cosmological data and will be a significant challenge for brute-force neural network-based analyses.

To test whether the notion of scaling laws extends to summary statistics beyond the power spectrum, and other loss functions, we briefly investigate the problem of neural posterior estimation with Kullback-Leibler loss, based on wavelet scattering coefficients.

The remainder of this paper is organized as follows. Section \ref{QuijoteSims} describes our simulation suite. Section \ref{InfoCont} presents methods for measuring parameter information content in cosmological datasets. Section \ref{NN} details our neural network methodology. In section \ref{convergence} we present our method for estimating the number of required training sets to asymptote to the Cramér-Rao information bound. In order to test the extrapolation of the resulting power law scaling predictions we present the Big Sobol Sequence (BSQ) in section \ref{sec:BSQ} comprising 32,768 simulations designed specifically for testing neural network convergence and quantifying information content. Our analysis includes careful consideration of the important topic of network architecture selection and hyperparameter optimization (see Appendix~\ref{archi-search}). Testing whether the finding of power law scaling translates to other summary statistics and loss functions, we find in Section \ref{sec:waveletStats} a scaling law for neural posterior estimation based on wavelet scattering transform coefficients with Kullback-Leibler loss function. We discuss our findings and conclude in section \ref{sec:Conclusions}.

\section{The Quijote simulation suite}\label{QuijoteSims}
The Quijote simulation suite is a collection of more than 88,000 full N-body dark matter simulations designed to probe the large-scale structure of the universe \citep{Villaescusa-Navarro:2019bje}. Using the TreePM code GADGET-III which is an improved version of GADGET-II \citep{Springel_2005}, these simulations span more than 40,000 cosmological models, exploring a parameter space defined by ${\Omega_m, \Omega_b, h, n_s, \sigma_8, M_\nu, w}$. Each simulation models the evolution of dark matter in a cubic volume of 1$(Gpc/h)^3$, computed at three spatial resolutions: $256^3$, $512^3$, and $1024^3$ grid points.

The suite includes several complementary subsets designed for different analysis approaches. For instance, the Latin Hypercube (LH) subset consists of 2,000 simulations sampling five parameters ${\Omega_m, \Omega_b, h, n_s, \sigma_8}$, specifically configured for machine learning applications. These parameters are sampled uniformly within ranges centered around the Planck 2018 best-fit values \citep{PlanckOverview2020}. 

Additionally, the Quijote suite provides a large set of simulations designed to facilitate the computation of Fisher matrices for arbitrary, informative statistics which we will discuss in section \ref{InfoCont}. This set contains 15,000  simulations at a fixed fiducial cosmology of $\Omega_m=0.3175,$ $\Omega_b=0.049,$ $h=0.6711,$ $n_s=0.96244$, and $\sigma_8=0.834$ and 500 pairs of simulations to support finite difference computations. These pairs are symmetrically displaced from the fiducial cosmology along each parameter direction. Each pair of simulations has identical initial conditions, suppressing sample variance in the finite difference computation.

For our investigation of simulation-based inference using neural networks, we focus on the matter power spectrum $P(k)$ in real space measured at redshift $z=0$. We analyze scales up to $k_{max}=0.807 h/\text{Mpc}$, corresponding to the $0.25$ times Nyquist frequency of a $1024^3$ grid in our 1$(Gpc/h)^3$ volume. The power spectrum is evaluated at multiples of the fundamental frequency, yielding a 128-dimensional measurement vector over this $k$ range. 

\section{Information content: Fisher information and forecasts}\label{InfoCont}
The Fisher information formalism \citep{1997ApJ...480...22T, Tegmark_1997, Alsing:2017var} provides a framework for quantifying the information about cosmological parameters $\theta$ contained in data $d(\theta)$. This approach yields insights into maximum likelihood parameter estimates and their uncertainties. For unbiased estimators, the Cramér-Rao information inequality \citep{Rao1945, Cramer1946} establishes a fundamental lower bound on parameter variance
\begin{equation}\label{infoineq}
    \text{Var}(\theta)\geq (F^{-1})_{\theta\theta},
\end{equation}
where $F$ is the Fisher information matrix, defined as:
$F=-\langle\nabla_\theta\nabla^T_\theta \mathcal{L}\rangle=\langle\nabla_\theta\mathcal{L}\nabla^T_\theta\mathcal{L}\rangle$, and $\mathcal{L}$ is the log-likelihood of the statistic. Here, the expectation value is taken over different realizations at fixed $\theta$, and the Fisher matrix in Eq.~\eqref{infoineq} is evaluated at the fiducial parameter value $\theta=\theta_\star$.

In the asymptotic limit of a large number of modes, and for informative statistics, the power spectrum likelihood is approximately Gaussian, so the log-likelihood takes the form
\begin{equation}
\mathcal{L}=-\frac{1}{2}(d-\mu(\theta))^T C^{-1}(d-\mu(\theta))-\frac{1}{2}\ln|2\pi C|,
\end{equation}
leading to a Fisher information matrix \citep{Alsing:2017var} of
\begin{equation}\label{FIM}
F=\nabla_{\theta}\mu^T C^{-1}\nabla_{\theta}\mu.
\end{equation}
The convergence of both $\nabla_\theta\mu$ and the covariance matrix $C$ in Eq.~\eqref{FIM} depends critically on the number of simulations \cite{coulton2023estimatefisherinformationmatrices}. Since the dark matter clustering statistics used in this paper are highly informative, we have found that the number of simulations provided in the Quijote suite (see Section \ref{QuijoteSims}) suffice for computing the finite difference approximation of the derivatives in Eq.~\eqref{FIM}.

At the maximum likelihood point
\begin{equation*}
\nabla_\theta \mathcal{L}|_{\text{ML}}=\nabla{\theta}\mu^T|_{\text{ML}} C^{-1}(d-\mu(\theta_{\text{ML}}))=0.
\end{equation*}
A Taylor expansion of this score function around the fiducial point $\theta_\star$ yields the Quasi-Maximum Likelihood estimator
\begin{equation}\label{QML}
\theta_\text{ML}=\theta_\star+F^{-1}\nabla_{\theta}\mu^T({\theta_\star}) C^{-1}(d-\mu(\theta_{\star}))
\end{equation}
at leading order in $\theta_\text{ML}-\theta_\star$. For parameters far from the fiducial point, Eq.~\eqref{QML} must be iterated until convergence, requiring covariance matrix evaluation at each intermediate point. However, this iteration is computationally prohibitive due to the large number of simulations required at each step.

In our analysis, we use the power spectrum $P(k)$ from the Quijote suite to establish optimal information bounds. The Fisher information matrix $F_{P(k)}$ is computed using the suite's 15,000 fiducial simulations for the covariance matrix $C$ and 500 pairs of finite difference simulations for the derivative terms $\nabla_\theta\mu$. This framework provides a benchmark against which we can evaluate the performance of neural network-based inference methods, which we discuss in the following section.

\section{Neural estimator}\label{NN}
Neural networks have become increasingly important tools in astrophysics and cosmology for extracting underlying parameters from simulated or observational data, owing to their ability to learn complex inverse models through chaining linear operations and nonlinear activation functions, combined with their differentiability, enabling efficient training with stochastic gradient descent. Developing effective models requires advances in multiple areas: architecture design, hyperparameter optimization, weight and bias initialization, and the relationship between network structure and the underlying physics of the data. In the following, we explore the application of neural networks in inferring cosmological parameters from power spectra $P(k)$ with particular attention to architecture selection, training, and hyperparameter tuning. 

\subsection*{Selection of the neural architecture}
\begin{figure}[htbp]
    \centering
    \includegraphics[width=3.in]{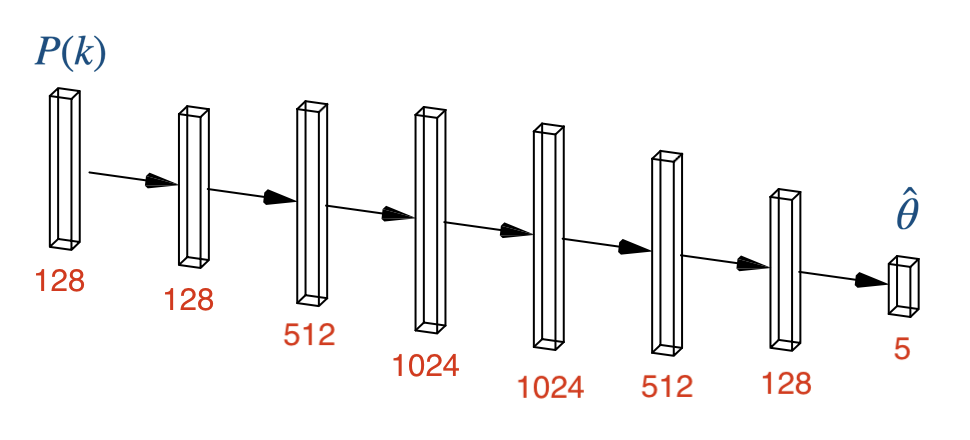}
    \caption{An illustration of \textsc{PowerSpectraNet} architecture. This infers cosmological parameters $\hat{\theta}:\{\Omega_m, \Omega_b, h, n_s, \sigma_8\}$ from the Power Spectrum $P(k)$. The dimension of each layer is mentioned at the bottom of the corresponding layer.}
    \label{PK3}
\end{figure}

Multi-layer Perceptrons (MLPs) demonstrate superior performance compared to linear regression models for nonlinear parameter estimation tasks. As fully connected feed-forward neural networks with multiple hidden layers and nonlinear activation functions, MLPs excel at capturing complex relationships in data. We employ an MLP to infer cosmological parameters from simulated power spectra $P(k)$. After extensive architecture exploration detailed in Appendix~\ref{archi-search}, we developed a model named \textsc{PowerSpectraNet} (Figure~\ref{PK3}) consisting of 6 hidden layers, each followed by LeakyReLU(0.5) activation. The hidden layer dimensions follow an expanding-contracting pattern $(128\rightarrow512\rightarrow1024\rightarrow1024\rightarrow512\rightarrow128)$, which improves parameter estimation performance. We will study the effect of variations in the architecture in Appendix~\ref{archi-search}.

For neural network training, we divide the 2000 Latin Hypercube simulations into training (1500), validation (300), and test (200) sets. To manage memory constraints during training, we process the data in batches of 16. All power spectra are normalized before being input into the model to improve convergence. We use the Adam optimizer with an initial learning rate of 0.001 and momentum ($\beta$) of 0.9 to minimize the log MSE loss $\mathcal{L}$ between the true parameter $\theta$ and inferred parameter $\hat{\theta}$
\begin{align}
\label{logloss}
\mathcal{L} 
&=\sum_i \ln \left(\frac{1}{N_{batch}}\sum_{n=0}^{N_{batch}} \big[\theta_i^{(n)}-\hat{\theta}_i^{(n)}\big]^2\right)
\end{align}
where $i$ denotes different parameter classes.
The learning rate is reduced by a factor $\gamma=0.9$ every 10 epochs using PyTorch's StepLR scheduler. These hyperparameters have been chosen after an extensive hyperparameter search, as shown in Appendix~\ref{archi-search}. Training the model on the 2000 Latin Hypercube $P(k)$ samples for 500 epochs required approximately one hour on a V100 GPU, with model weights at the lowest validation loss retained for inference.

\section{Convergence analysis and scaling law to estimate simulation requirements}\label{convergence}
\begin{figure}[htbp]
    \centering
    \includegraphics[width=3.4in]{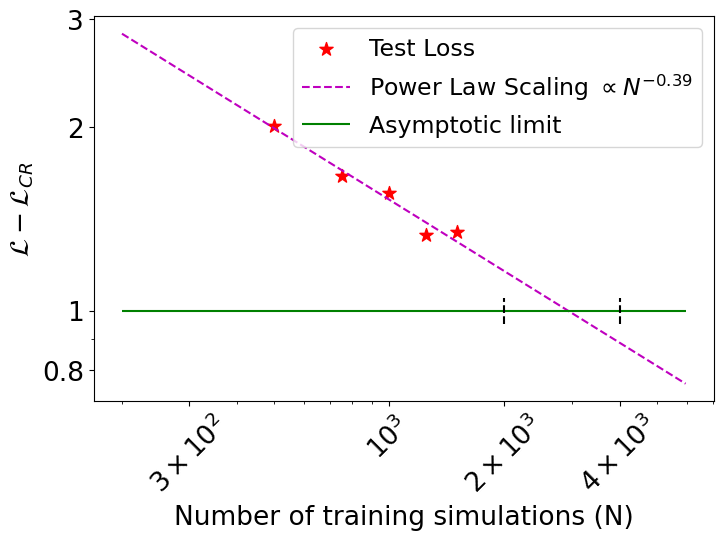}
    \caption{Loss (Eq.~\ref{logloss}) on held out test data as a function of number of Quijote LH simulations. The loss  asymptotes to the Cramér-Rao (C-R) bound $\mathcal{L}_{CR}$ by power law decay $\propto N^{-0.39}$ (cf. Eq.~\ref{scaling}). We mark on the abcissa the asymptotic regime, defined as where the test loss becomes $e$ times the Cramèr-Rao bound. Based on our initial training set of 1500  we predict that the information extracted by the neural summary is not yet optimal and that several thousand simulations are needed to be in the asymptotic regime. This prediction is verified in Figure~\ref{corner_BSQ}.}
    \label{scaling_LH}
\end{figure}

To understand how many simulations are required for optimal parameter inference, we systematically study how network performance scales with training set size. After training, we use the neural compression $\hat{\theta}$ from fiducial and finite difference simulations to evaluate Fisher information from the neural summary, comparing it with the total information $F_{P(k)}$ of the original data. Figure \ref{scaling_LH} shows the test loss as a function of training simulation count, where the optimal performance is bounded below by
\begin{equation}\label{CR_bound}
\mathcal{L}_{CR}=\text{Tr}[\ln(\mathrm{diag}(F_{P(k)}^{-1}))].
\end{equation}

Scaling laws have been found to be ubiquitous in deep learning systems \citep{hestness2017deeplearningscalingpredictable} including in recent, seminal scaling studies of transformer models \citep{kaplan2020scaling,edwards2024scalinglaws}. Inspired by these studies, and using the fact that the information in finite, noisy data sets is ultimately bounded, we hypothesize that the relationship follows a power law that asymptotes with the increasing number of simulations $N$
\begin{equation}\label{scaling}
\mathcal{L}=\mathcal{L}_{CR}+cN^{-\alpha},
\end{equation}
where $c$ is a constant and the power law fitting to the test losses yields $\alpha \approx 0.39$.
The figure reveals that the test log-loss scales as a power law of the number of simulations. The 1500  training simulations available in the Latin Hypercube set are insufficient to reach the asymptotic regime defined here as where the test loss becomes a factor of $e$ times the Cramèr-Rao bound. The power law scaling predicts that several thousand more simulations are needed to saturate the information content of the non-linear matter $P(k)$.

\begin{figure}[htbp]
    \centering
    \includegraphics[width=3.4in]{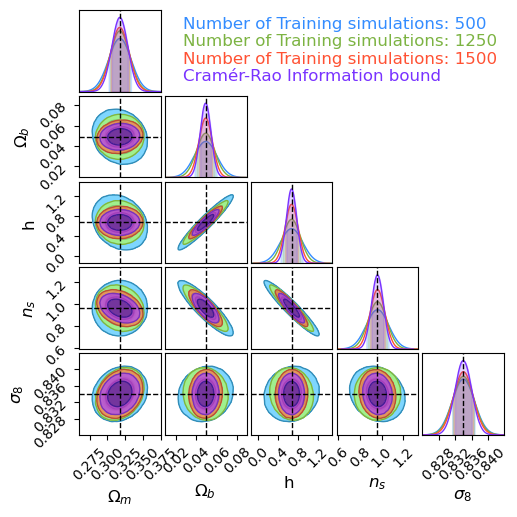}
    \caption{Fisher information in neural summaries vs optimal $P(k)$. NNs trained on 500, 1250, and 1500 LH simulations show increasingly tighter parameter constraints but do not saturate the information bound of an optimal estimator based on the $P(k)$.}
    \label{corner_LH}
\end{figure}

Although decreasing log MSE loss does not guarantee increased Fisher information across all parameters, we expect information gains for at least some parameters. In the ideal case, where loss contributions from each parameter are comparable, uncertainties should decrease uniformly across all parameters until information saturation. Figure {\ref{corner_LH}} demonstrates the fraction of optimal information extracted by neural summaries at the fiducial parameter point, when trained on different numbers of simulations. More specifically, the parameter uncertainties are derived from the Fisher information matrix based on the neural summaries, calculated using \eqref{FIM}.

We will now describe the simulation suite we ran and used to test our prediction.

\subsection{The Big Sobol Sequence (BSQ) simulations}\label{sec:BSQ}
Our analysis of the Quijote Latin Hypercube simulations indicates the need for a larger training set to saturate the available cosmological information. To verify our predictions about the required simulation count and test convergence behavior, we generated a new suite of 32,768 N-body simulations using Gadget-III. We call these simulations the \textsc{Big Sobol Sequence} (BSQ) and make them publicly available\footnote{\url{https://quijote-simulations.readthedocs.io/en/latest/bsq.html}}. 
While  matched to the original simulations in terms of particle number and volume, the new simulations were run with slightly more stringent force accuracy parameters: PMGRID=2048, ASMTH=3.0, RCUT=6.0, and
MaxSizeTimestep=0.005. All other n-body code parameters are identical to the settings in the original Quijote suite.

The simulations cover the same parameter space as the Quijote LH simulations, sampling the five parameters ${\Omega_m, \Omega_b, h, n_s, \sigma_8}$ uniformly within ranges centered around the Planck 2018 best-fit values \citep{PlanckOverview2020}. The BSQ simulations are generated using a Sobol sequence design \citep{Sobol1967OnTD} rather than a Latin Hypercube design. 
We utilize 16,600 of these simulations, maintaining the same 15:3:2 ratio for train, validation, and test sets as in our previous analysis. We have verified that each parameter (and each parameter pair) is sampled uniformly across the parameter range. To ensure comparability, we employ the same \textsc{PowerSpectraNet}  architecture shown in Figure \ref{PK3} and hyperparameters that are identical to the ones used for the Latin Hypercube simulations.

\begin{figure}[htbp]
    \centering
    \includegraphics[width=3.4in]{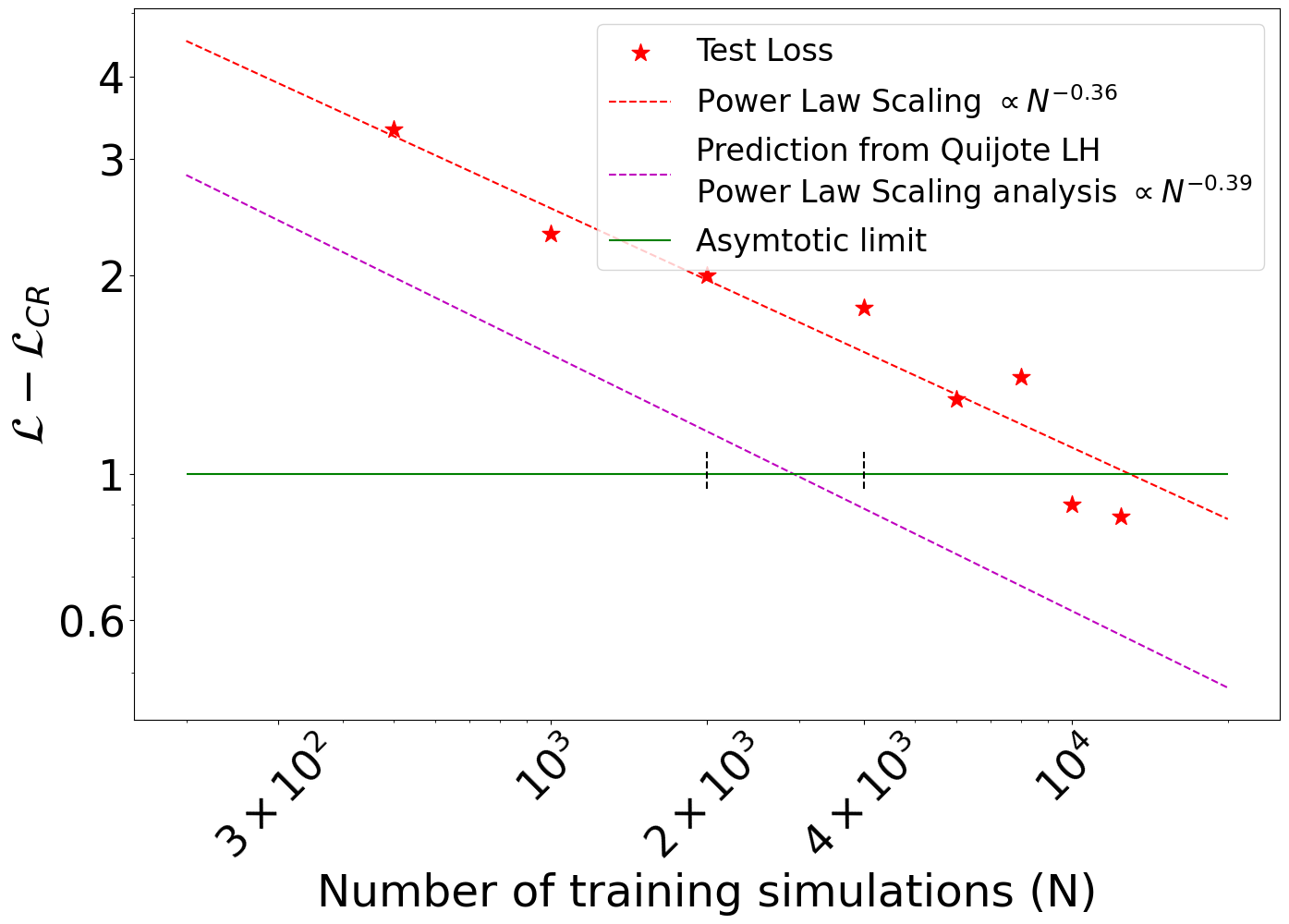}
    \caption{Test loss \eqref{logloss} as a function of number of BSQ simulations. The loss follows a simple power law scaling law across 2 orders of magnitude with a power law slope similar to the LH simulations (shown for comparison). The combined test loss for all parameters does not reach the optimal Fisher information computed at the fiducial point. The test loss is computed over the full prior range in parameter space rather than at a single point in the center of the training data. The loss of BSQ simulations are slightly higher than the LH simulations, but Figure {\ref{corner_BSQ}} shows that the NN achieves asymptotic optimality at the fiducial point at 4000 simulations.}
    \label{scaling_BSQ}
\end{figure}

Figure \ref{scaling_BSQ} shows the relationship between loss and the number of simulations, demonstrating that with sufficient training data, our neural network achieves optimal summary statistics as the loss curve approaches the theoretical information bound. Networks trained on the BSQ simulations exhibit a slightly higher test loss compared to the LH simulations for the same size of training set. The slightly improved accuracy of the BSQ simulations is unlikely to produce a noticeable effect on the power spectra in the $k$ range used for this study. A more likely reason is the different sampling of the parameter space of the Sobol sequence and the Latin Hypercube designs. This difference in sampling could lead to variations in the effective training set size and the resulting model performance. However, the test loss of BSQ simulations does also follow a power law scaling of $N^{-0.36}$, with a power law exponent that is within 10\% of the exponent we found earlier from the Latin Hypercube simulations.

\begin{figure}[htbp]
    \centering
    \includegraphics[width=3.4in]{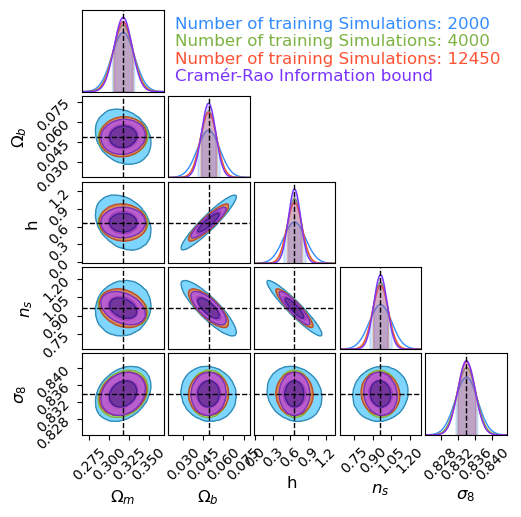}
    \caption{Fisher information in neural summaries vs optimal $P(k)$. Information starts saturating near 4000 BSQ simulations and the corresponding neural summary is nearly as informative as the optimal $P(k)$.}
    \label{corner_BSQ}
\end{figure}

The parameter uncertainties inferred from models trained on different numbers of simulations appear in Figure {\ref{corner_BSQ}}. These results demonstrate that the information content saturates for nearly 4,000 simulations, with minimal improvement beyond this point. This is consistent with our prediction based on the power law scaling analysis of the Latin Hypercube simulations.

The Fisher matrix results seem to be better than expected based on the higher test loss in the BSQ suite. Based on the offset of the scaling laws we may have expected that more, perhaps 10,000 simulations would have been necessary to reach the asymptotic regime. Consider that the Fisher information is computed at the fiducial model in the center of the parameter space while the test loss is an average over the full prior volume. The Sobol sequence is strongly self-avoiding and so there will be models in the test set that are near the boundaries of the prior space, where the feature extraction may become less efficient. In summary, we find that the difference in parameter sampling do affect the constant in the test loss scaling; the two sampling designs seem to provide similar Fisher information for similar training set size at the fiducial parameter set in the center of the prior volume.


\subsection{Robustness of scaling laws: Wavelet statistics}
\label{sec:waveletStats}
So far, we collected evidence for scaling laws that for feature extraction from the power spectrum using a quadratic loss function. In this subsection, we will go beyond this and look for scaling laws in a simulation based inference task based on Wavelet Scattering Transform (WST) coefficients \citep{Valogiannis_2022, eickenberg2022waveletmomentscosmologicalparameter, SimBIG:2023gke, DES:2023qwe}.

We have computed our wavelet statistics using the Kymatio package \citep{andreux2022kymatioscatteringtransformspython} from Quijote dark matter simulations of $128^3$ down-sampled resolution. To reduce the dynamic range we take the natural log of the WST coefficients as the summary vector. We find that some   WST coefficients are highly correlated in the fiducial simulations; this causes difficulties downstream, since such correlations render the covariance matrix nearly singular. This in turn leads to numerical instabilities in the Fisher matrix computation. To address this, we find highly correlated pairs  (those with Pearson's $|r|>0.99$ \citep{kendall1977advanced}) and then eliminate redundant WST coefficients from the parameter vector.

\begin{figure}[htbp]
    \centering
    \includegraphics[width=3.4in]
    {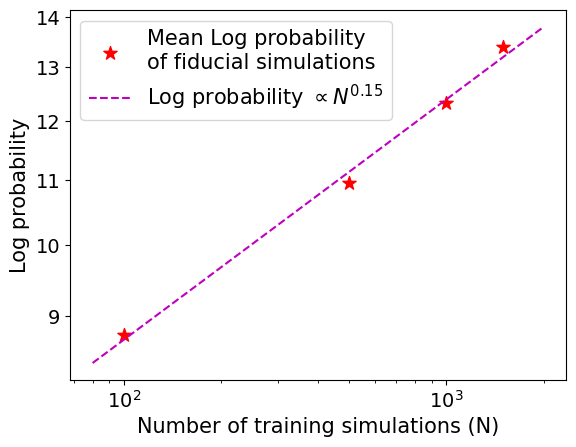}    \caption{Log-probability from an ensemble of NN trained on wavelets for different number of Quijote LH training simulations. The log-probability simply follows a power law similar to the $P(k)$ analysis.}
    \label{scaling_wst}
\end{figure}

We use Learning the Universe Implicit Likelihood Inference (LtU-ILI) \citep{Ho_2024} with the \texttt{lampe} backend to train an ensemble of neural posterior estimators (NPEs) consisting of Masked Autoregressive Flow (MAF) \citep{papamakarios2018maskedautoregressiveflowdensity}, Gaussianization Flow (gf) \citep{meng2020gaussianizationflows}, Nonlinear Independent Components Estimation (NICE) \citep{dinh2015nicenonlinearindependentcomponents}, Neural Spline Flow (NSF) \citep{durkan2019neuralsplineflows}, and Mixture Density Network (MDN) \citep{bishop1994mixture} on the wavelets and estimate the log-probability of the fiducial simulations. We repeat this process for each number of training simulations. The results suggest that the log-probability again follows a scaling law as the number of training simulations changes, as shown in Figure \ref{scaling_wst}.

\begin{figure}[htbp]
    \centering
    \includegraphics[width=3.5in]
    {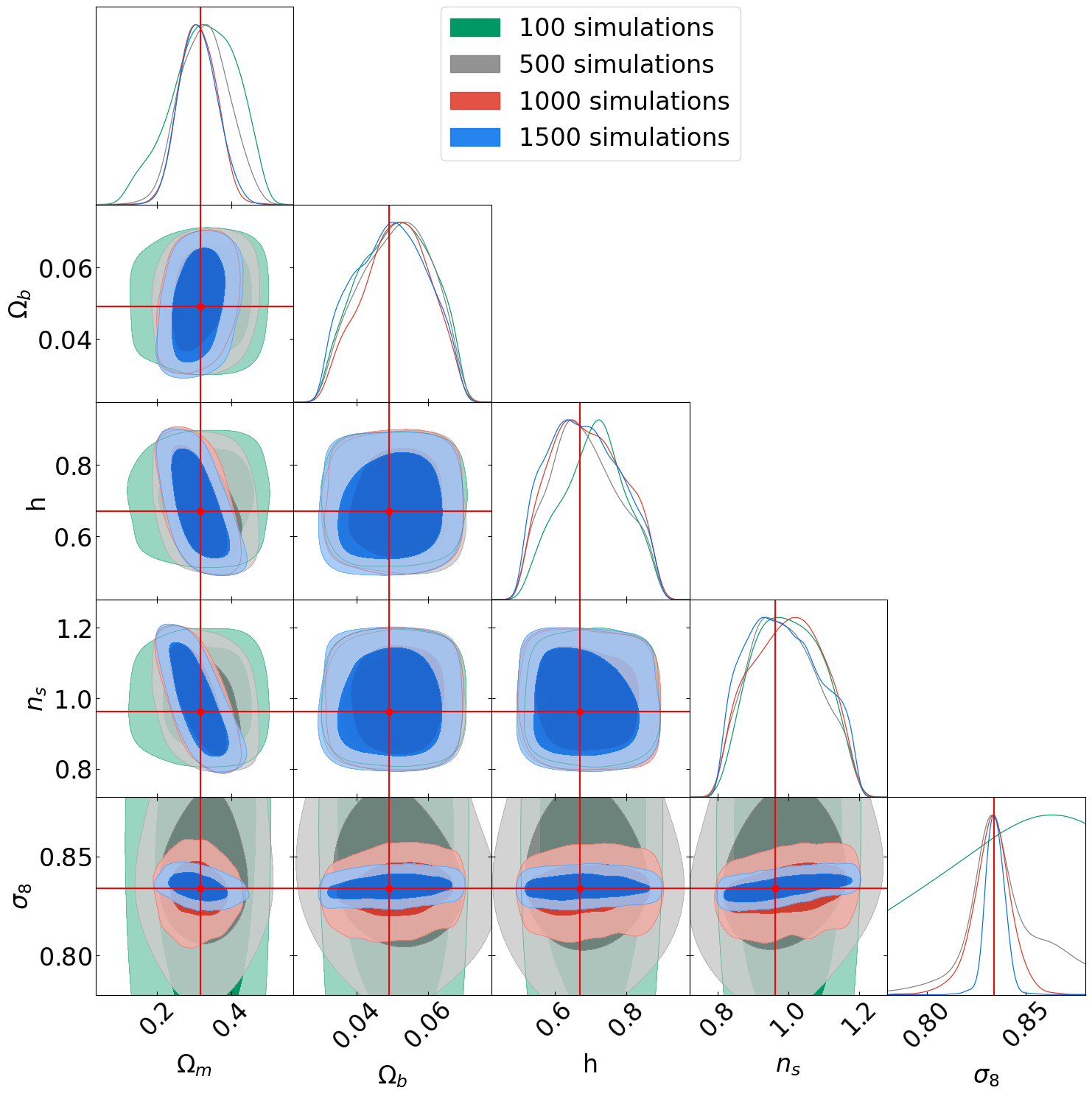}
    \caption{Scaling of information content of the neural posterior estimation based on wavelet summaries. The power law scaling of the log probability loss with the increasing number of simulations is reflected in improved constraints on $\Omega_m$ and $\sigma_8$. Other parameters show mild improvement.}
    \label{corner_wst}
\end{figure}

We draw 10000 samples from the learned neural posteriors for a simulation at the fiducial cosmology and use those samples to demonstrate the posterior distribution in Figure \ref{corner_wst} based on training with 100, 500, 1000 and 1500 simulations. The corner plot shows that the constraints on $\Omega_m$ and $\sigma_8$ decrease significantly with the increasing number of simulations, while other parameters show mild improvement with the number of training simulations.

\section{Discussion and conclusion}\label{sec:Conclusions}
This paper addresses a fundamental question in simulation-based inference for cosmology: How many simulations are required for neural networks to achieve optimal feature extraction for parameter inference? Our investigation of this question yields several important insights.

First, we demonstrate that current simulation suites, while impressive in scope, may be insufficient for optimal training of neural feature extractors. Specifically, we show that the 2,000 Latin Hypercube simulations in the Quijote suite, while valuable, do not provide enough training data for neural networks to reach the Cramér-Rao information bound, even for the relatively simple case of the matter power spectrum $P(k)$.

Through careful analysis of neural network performance scaling, we develop a quantitative framework for predicting simulation requirements. Our initial study using Latin Hypercube simulations suggests approximately 4,000 training simulations are needed to saturate the information content from $P(k)$, based on observed convergence patterns and a simple power-law scaling assumption.

This prediction motivated the creation of our \textsc{Big Sobol Sequence} (BSQ)\footnote{\url{https://quijote-simulations.readthedocs.io/en/latest/bsq.html}}, a new and publicly available suite of 32,768 N-body simulations that allows us to verify these scaling relationships and provide a comprehensive resource for future studies.

Analysis of networks trained on the BSQ simulations confirms our earlier predictions. The test loss follows a power law decay similar to that observed in the Latin Hypercube simulations, with information content indeed beginning to saturate around 4,000 simulations. While the sample efficiency of the two simulation designs differs, producing different constant factors for the scaling laws,  each simulation suite exhibited predictable power law scaling of the test loss with the size of the training set. 

Extending the analysis beyond the power spectrum and squared error loss, we ran a similar scaling study for neural posterior estimation of cosmological parameters from wavelet scattering transform coefficients. We find that the test Kullback-Leibler loss for the neural density estimation task also scales as a power law of training set size.

Our architectural studies reveal an important relationship between network depth and training set size. While deeper networks can potentially extract more information, we find that increasing the number of training simulations often provides greater improvement than adding network layers. 

These findings have significant implications for cosmological analysis. As we move toward more complex summary statistics and combine multiple probes, the simulation requirements will present a challenge save for innovations in training methodology or fast-forward models to accelerate the production of simulated training sets. Our results suggest that future simulation campaigns should carefully consider training set size requirements when allocating computational resources. 

Looking forward, several questions merit further investigation. The power law scaling we observe may have deeper theoretical foundations that could inform future network design. The difference we observe in sampling efficiency between the Sobol sequence and the Latin Hypercube sampling designs suggests that surveying other strategies to sample parameters for training sets may lead to more simulation efficient information extraction.  Additionally, extending this analysis to other summary statistics and their combinations could provide valuable insights for survey analysis. 
The BSQ simulation suite, which we make publicly available, provides a robust foundation for such investigations. 

Our study also motivates research into more simulation efficient ways to train neural networks for feature extraction. A recent study reports that in the case of limited training data, machine learning inference on weak lensing data outperforms when the NN training is conditioned on some complementary summary i.e. power spectrum $P(k)$ \citep{makinen2024hybridsummarystatistics}. The steeper scaling law we found in our case study of neural posterior estimation compared to estimating the feature extraction using regression of the posterior mean also suggests that some inference approaches may be more sample efficient than others. In the interest of reducing the simulation cost to train feature extractors, we plan to investigate hierarchical approaches that avoid training networks that ingest the full survey volume at full resolution all at once. 

As we prepare for next-generation surveys with larger volumes and higher galaxy number density, understanding the requirements for neural network performance becomes increasingly crucial. Our work provides both practical guidance for simulation campaigns and a framework for guiding neural information extraction in cosmological parameter inference.

\begin{acknowledgements}
AB acknowledges support from the Simons Foundation as part of the Simons Collaboration on Learning the Universe. The Flatiron Institute is supported by the Simons Foundation. All other post-analysis and neural training have been done on IAP's Infinity cluster; the authors acknowledge St\'ephane Rouberol for his efficient management of this facility. The BSQ simulations were generated using the Rusty cluster at the Flatiron Institute. The work of FVN is supported by the Simons Foundation. BDW acknowledges support from the DIM ORIGINES 2023 INFINITY NEXT grant. 
\end{acknowledgements}

\bibliographystyle{aa} 
\bibliography{aa54602-25} 
\newpage

\begin{appendix}
\section{Effect of varying the neural network architecture}\label{archi-search}
\begin{figure}[htbp]
    \centering
    \includegraphics[width=3.in]{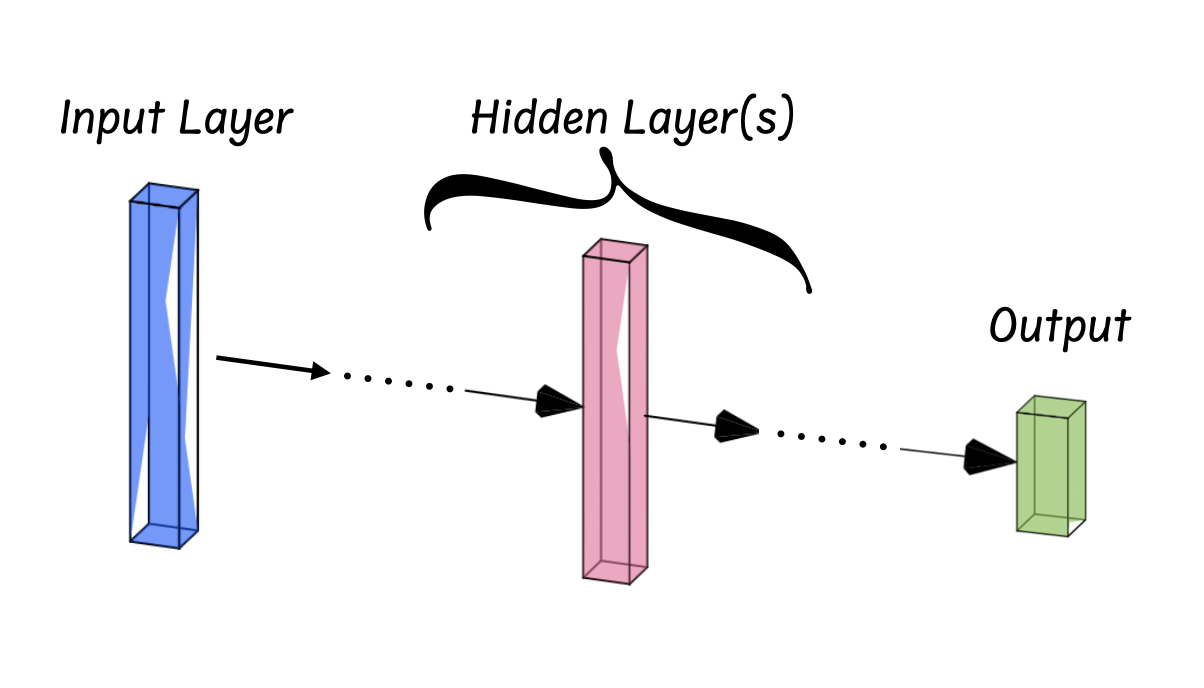}
    \caption{A generic MLP architecture of varying numbers of hidden layers of dimension 128. This is used to analyze the impact of the depth of the NN on the inference.}
    \label{MLP-generic}
\end{figure}

Neural network architecture plays a vital role in extracting information from data. While no straightforward method exists for identifying the optimal architecture, systematic exploration across different numbers of trainable parameters provides insight into the network depth and width required to obtain near-optimal results. We investigate this by examining a generic MLP architecture that varies the number of hidden layers while maintaining fixed input and output dimensions (Figure \ref{MLP-generic}). 

\begin{figure}[htbp]
    \centering
    \includegraphics[width=3.4in]{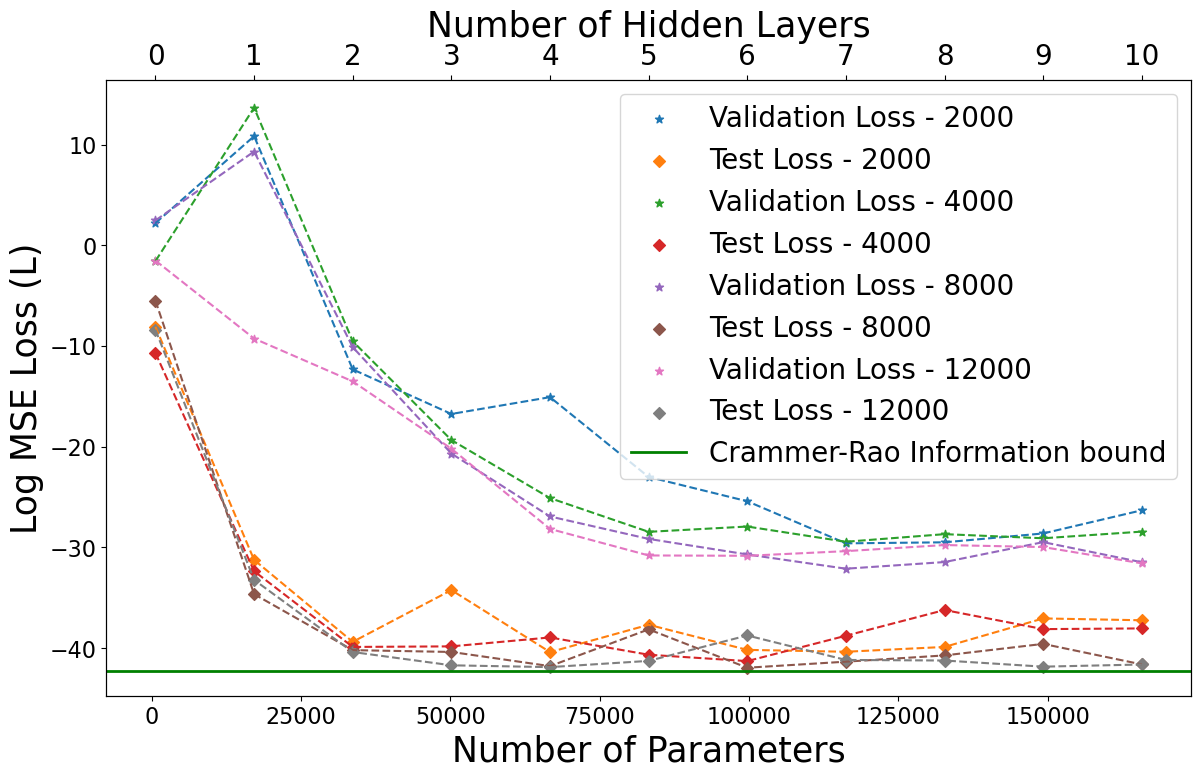}
    \caption{Loss vs depth of the MLP (i.e. number of trainable parameters). Shallower models require more realizations to reach the same performance.}
    \label{lossvsdepth}
\end{figure}

Given our task of inferring five cosmological parameters from $P(k)$, we set the input dimension to 128 and the output dimension to 5. Each hidden layer maintains the same dimension as the input layer, and we employ LeakyReLU(0.5) activation throughout.
Figure \ref{lossvsdepth} presents validation and test losses for MLPs with hidden layers ranging from 0 to 10, trained on varying numbers of simulations. Our analysis reveals that similar levels of Fisher information can be achieved either by increasing the training set size or by deepening the network. However, simply increasing the number of trainable parameters does not guarantee improved information extraction, since we find that deeper architectures can make optimization more challenging.
\\

\subsection*{Development of \textsc{PowerSpectraNet}}
\begin{figure}[htbp]
    \centering
    \includegraphics[width=3.4in]{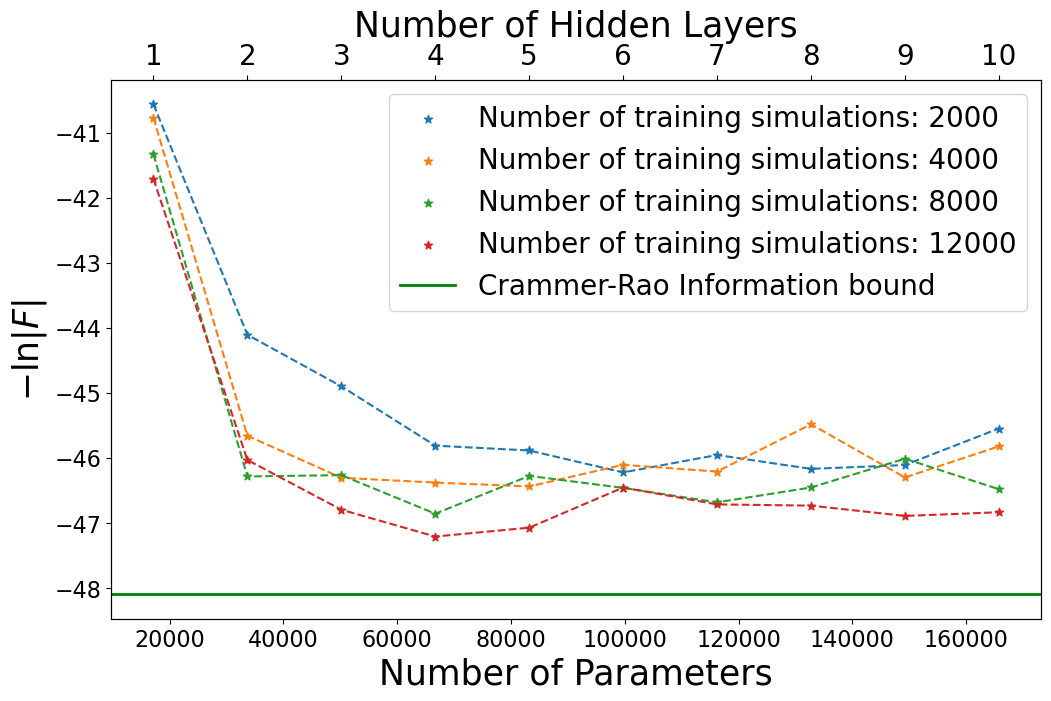}
    \caption{Fisher information as a function of the depth of the MLP (i.e. number of trainable parameters). Information increases when the MLP is trained on enough number of simulations. Information saturates at 6 hidden layers for 2000 realizations and at this point, information does not change much even when the number of simulations is changed.}
    \label{lnF}
\end{figure}

Our extensive architecture search revealed that even for the relatively simple $P(k)$ summary statistic, optimal performance requires seven fully connected layers (six hidden layers plus output) with LeakyReLU activation (Figure {\ref{PK3}}). We evaluated numerous MLPs of varying depth using 2,000 BSQ $P(k)$ samples, finding that performance plateaus at six hidden layers, with no additional information gain from deeper architectures (Figure \ref{lnF}). This pattern holds consistently across different training set sizes, providing two key justifications for our architectural choice. Further refinement showed improved performance when systematically varying layer widths in an expanding-contracting pattern, leading to the final \textsc{PowerSpectraNet} architecture.

\subsection*{Hyperparameter selection}\label{hyperparameter-tuning}
Given the computational cost of hyperparameter optimization for deep architectures, we conducted an extensive hyperparameter search using a single-layer perceptron (equivalent to linear regression) on $P(k)$. This approach aligns with recent studies of large neural networks \citep{kaplan2020scaling}, which suggest hyperparameter scalability across network depths. 
We utilize the Weights and Biases \citep{wandb} tool to optimize key hyperparameters for our neural network. Our search explores batch sizes in the range $[16, 64]$, activation functions (tanh, ReLU, sigmoid, and a custom activation function (see appendix \ref{Activation})), optimizers (SGD, Adam, and BFGS), and learning rate schedulers (CyclicLR and StepLR). Additionally, we examine learning rates within $[0, 0.01]$ and tune momentum and gamma values in the range $[0.1, 1]$. This study aims to identify the optimal combination of hyperparameters to enhance model performance effectively. The optimal hyperparameter values appear in Table \ref{hyperparamtable}.
\begin{table}[htbp]
        \caption{Hyperparameters are optimized for the linear model and kept fixed throughout the paper while testing different architectures.}
      \centering
      \begin{tabular}{l r}
        \toprule
        Hyperparameter & Value\\
        \toprule
        Batchsize & 16 \\
        Nonlinear Activation & LeakyReLU(0.5)\\
        Optimizer & Adam \\
        Momentum & 0.9 \\
        Learning Rate Scheduler & StepLR \\
        Initial Learning Rate & 0.001 \\        
        Stepsize & 10\\
        Gamma & 0.9 \\
        \toprule
      \end{tabular}
      
      \label{hyperparamtable}
\end{table}

\section{Activation functions}\label{Activation}
In this section, we describe custom activation functions considered in our study other than the traditional ones to find out the best possible set of hyperparameters for neural network training. Activation functions introduce non-linearity into neural networks, allowing them to learn complex patterns. Below are the key functions used in this paper. These are piecewise continuous nonlinear functions designed to provide smooth transitions across different input ranges while mitigating issues like vanishing gradients and dead neurons. The functions are linear for $|x|>1$ and follow a cubic polynomial in the central region, ensuring smoothness and differentiability for all   $x\in\mathbb{R}$.

\subsection*{SymmCubic}
\begin{figure}[htbp]
    \centering
    \includegraphics[width=3.2in]{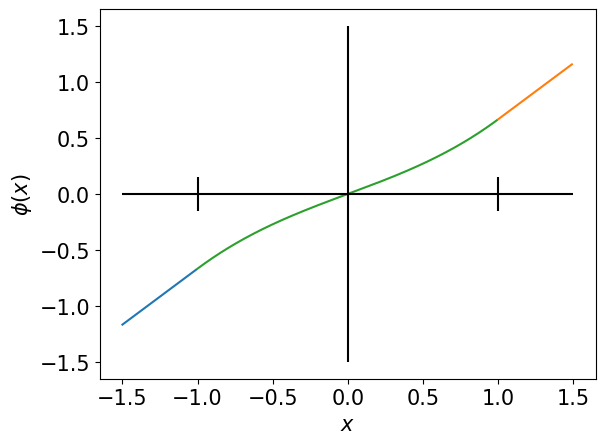}
    \caption{SymmCubic}
    \label{SymmCubic}
\end{figure}

This activation function has a cubic nonlinearity and it has symmetry $\phi(-x)=-\phi(x)$ (Figure \ref{SymmCubic}). 
\begin{equation}
\phi(x)=
    \begin{cases}    
        x+1/3 & x<-1\\
        x(x^2+3)/6 & -1\le x \le 1\\
        x-1/3 & x>1  
    \end{cases} 
\end{equation}

\subsection*{LeakyCubic}
\begin{figure}[htbp]
    \centering
    \includegraphics[width=3.2in]{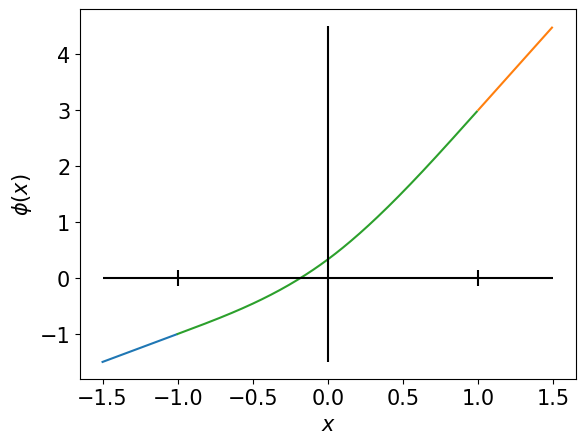}
    \caption{LeakyCubic}
    \label{LeakyCubic}
\end{figure}

This activation function resembles the leaky ReLU but replaces the kink at $x=0$ with a smooth cubic piece (Figure \ref{LeakyCubic}).

\begin{equation}
\phi(x)=
    \begin{cases}
        x & x<-1\\
        -|x|^3/3+x(x+2)+1/3 & -1\le x \le 1\\
        3x & x>1        
    \end{cases}
\end{equation}

\end{appendix}

\end{document}